\documentclass{article}
\usepackage[margin=2.5cm]{geometry}
\usepackage{hyperref}
\usepackage{natbib}
\usepackage{amsthm}

\usepackage{amssymb}
\usepackage{epstopdf}
\usepackage{mathtools}
\usepackage{amsfonts}
\usepackage{amsmath}
\usepackage{graphicx}
 \usepackage{booktabs}
\usepackage{algorithm}
\usepackage[table,xcdraw]{xcolor}
\usepackage{algpseudocode}

\newtheorem{prop}{Property}

\newcommand{\TODO}[1]{\textcolor{red}{\bf[TODO: #1]}}
\newcommand{\OMIT}[1]{}
\newcommand{\RR}{\mathbb{R}}    %reals

\begin{document}
%\firstpage{1}
%\subtitle{Studies of Phenotypes and Clinical Applications}

%\title[DropLasso: A robust variant of Lasso for single cell RNA-seq data]{DropLasso: A robust variant of Lasso for single cell RNA-seq data}
%\author[Khalfaoui and Vert]{Beyrem Khalfaoui\,$^{\text{\sfb 1,2}}$ and Jean-Philippe Vert\,$^{\text{\sfb 1,2,3,}*}$ }
\title{DropLasso: A robust variant of Lasso for single cell RNA-seq data}
\author{Beyrem Khalfaoui\,$^{\text{1,2}}$ and Jean-Philippe Vert\,$^{\text{1,2,3,}}$\\ \\
$^{\text{\sf 1}}$  MINES ParisTech, PSL Research University, \\CBIO - Centre for Computational Biology, F-75006 Paris, France\\
$^{\text{\sf 2}}$ Institut Curie, PSL Research University, INSERM, U900, F-75005 Paris, France.\\  
$^{\text{\sf 3}}$ Ecole Normale Sup\'erieure, Department of Mathematics and Applications, \\CNRS, PSL Research University, F-75005 Paris, France.\\
\url{firstname.lastname@mines-paristech.fr}}

%\corresp{$^\ast$To whom correspondence should be addressed.}

%\history{Received on XXXXX; revised on XXXXX; accepted on XXXXX}

%\editor{Associate Editor: XXXXXXX}

\date{}
\maketitle

\begin{abstract}
Single-cell RNA sequencing (scRNA-seq) is a fast growing approach to measure the genome-wide transcriptome of many individual cells in parallel, but results in noisy data with many dropout events. Existing methods to learn molecular signatures from bulk transcriptomic data may therefore not be adapted to scRNA-seq data, in order to automatically classify individual cells into predefined classes.\\
We propose a new method called DropLasso to learn a molecular signature from scRNA-seq data. DropLasso extends the dropout regularisation technique, popular in neural network training, to estimate sparse linear models. It is well adapted to data corrupted by dropout noise, such as scRNA-seq data, and we clarify how it relates to elastic net regularisation. We provide promising results on simulated and real scRNA-seq data, suggesting that DropLasso may be better adapted than standard regularisations to infer molecular signatures from scRNA-seq data.\\
%\textbf{Availability:} 
DropLasso is freely available as an R package at \url{https://github.com/jpvert/droplasso}\\
%\textbf{Contact:} \href{jean-philippe.vert@mines-paristech.fr}{jean-philippe.vert@mines-paristech.fr}\\
%\textbf{Supplementary information:} Supplementary data are available at \textit{Bioinformatics} online.}
\end{abstract}

\section{Introduction}
The fast paced development of massively parallel sequencing technologies and protocols has made it possible to measure gene expression with more precision and less cost in recent years. Single-cell RNA sequencing (scRNA-seq), in particular, is a fast growing approach to measure the genome-wide transcriptome of many individual cells in parallel \citep{Kolodziejczyk2015technology}. By giving access to cell-to-cell variability, it represents a major advance compared to standard ``bulk'' RNA sequencing to investigate complex heterogeneous tissues \citep{Macosko2015Highly,Tasic2016Adult,Zeisel2015Cell,Villani2017Single} and study dynamic biological processes such as embryo development \citep{Deng2014Single} and cancer \citep{Patel2014Single}.

The analysis of scRNA-seq data is however challenging and raises a number of specific modelling and computational issues \citep{Ozsolak2011RNA,Bacher2016Design}. In particular, since a tiny amount of RNA is present in each cell, a large fraction of polyadenylated RNA can be stochastically lost during sample preparation steps including cell lysis, reverse transcription or amplification. As a result, many genes fail to be detected even though they are expressed, a type of errors usually referred to as \emph{dropouts}. In a standard scRNA-seq experiment it is common to observe more than 80\% of genes with no apparent expression in each single cell, an important proportion of which are in fact dropout errors \citep{Kharchenko2014Bayesian}. The presence of so many zeros in the raw data can have significant impact on the downstream analysis and biological conclusions, and has given rise to new statistical models for data normalisation and visualisation \citep{Pierson2015Dimensionality,Risso2018ZINB} or gene differential analysis \citep{Kharchenko2014Bayesian}.

Besides exploratory analysis and gene-per-gene differential analysis, a promising use of scRNA-seq technology is to automatically classify individual cells into pre-specified classes, such as particular cell types in a cancer tissue. This requires to establish cell type specific ``molecular signatures'' that could be shared and used consistently across laboratories, just like standard molecular signatures are commonly used to classify tumour samples into subtypes from bulk transcriptomic data \citep{Ramaswamy2001Multiclass,Soerlie2001Gene,Sorlie2003Repeated}. From a methodological point of view, molecular signatures are based on a \emph{supervised analysis}, where a model is trained to associate each genome-wide transcriptomic profile to a particular class, using a set of profiles with class annotation to select the genes in the signature and fit the parameters of the models. While the classes themselves may be the result of an unsupervised analysis, just like breast cancer subtypes which were initially defined from a first unsupervised clustering analysis of a set of tumours \citep{Perou2000Molecular}, the development of a signature to classify any new sample into one of the classes is generally based on a method for supervised classification or regression.

Signatures based on a few selected genes, such as the 70-gene signature for breast cancer prognosis of \citet{Vijver2002gene-expression}, are particularly useful both for interpretability of the signature, and to limit the risk of overfitting the training set. Many techniques exist to train molecular signatures on bulk transcriptomic data  \citep{Haury2010stability}, however, they may not be adapted to scRNA-seq data due to the inflation of zeros resulting from dropout events.

Interestingly and independently, the term ``dropout'' has also gained popularity in the machine learning community in recent years, as a powerful technique to regularise deep neural networks \citep{Srivastava2014Dropout}. Dropout regularisation works by randomly removing connexions or nodes during parameter optimisation of a neural network. On a simple linear model (a.k.a. single-layer neural network), this is equivalent to randomly creating some dropout noise to the training examples, i.e., to randomly set some features to zeros in the training examples \citep{Wager2013Dropout,Baldi2013Understanding}. Several explanations have been proposed for the empirical success of dropout regularisation. \citet{Srivastava2014Dropout} motivated the technique as a way to perform an ensemble average of many neural networks, likely to reduce the generalisation error by reducing the variance of the estimator, similar to other ensemble averaging techniques like bagging \citep{Breiman1996Bagging} or random forests \citep{Breiman2001Random}. Another justification for the relevance of dropout regularisation, particularly in the linear model case, is that it performs an intrinsic data-dependent regularisation of the estimator \citep{Wager2013Dropout,Baldi2013Understanding} which is particularly interesting in the presence of rare but important features. Yet another justification for dropout regularisation, particularly relevant for us, is that it can be interpreted as a \emph{data augmentation} technique, a general method that amounts to adding virtual training examples by applying some transformation to the actual training examples, such as rotations of images or corruption by some Gaussian noise; the hypothesis being that the class should not change after transformation. Data augmentation has a long history in machine learning \citep[e.g.,][]{Scholkopf1996Incorporating}, and is a key ingredient of many modern successful applications of machine learning such as image classification \citep{Krizhevsky2012ImageNet}. As shown by \citet{Maaten2013Learning}, dropout regularisation in the linear model case can be interpreted as a data augmentation technique, where corruption by dropout noise enforces the model to be robust to dropout events in the test data, e.g., to blanking of some pixels on images or to removal of some words in a document. \citet{Wager2014Altitude} show that in some cases, data augmentation with dropout noise allows to train model that should be insensitive to such noise more efficiently than without.

Since scRNA-seq data are inherently corrupted by dropout noise, we therefore propose that dropout regularisation may be a sound approach to make the predictive model robust to this form of noise, and consequently to improve their generalisation performance on scRNA-seq supervised classification. Since plain dropout regularisation does not lead to feature selection and to the identification of a limited number of genes to form a molecular signature, we furthermore propose an extension of dropout regularisation, which we call \emph{DropLasso} regularisation, obtained by adding a sparsity-inducing $\ell_1$ regularisation to the objective function of the dropout regularisation, just like \emph{lasso} regression adds an $\ell_1$ penalty to a mean squared error criterion in order to estimate a sparse model \citep{Tibshirani1996Regression}. We show that the $\ell_1$ penalty can be integrated in the standard stochastic gradient algorithm used to implement dropout regularisation, resulting in a scalable stochastic \emph{proximal} gradient descent formulation of DropLasso. We also clarify the regularisation property of DropLasso, and show that it is to elastic net regularisation what plain dropout regularisation is to the plain ridge regularisation. Finally, we provide promising results on simulated and real scRNA-seq data, suggesting that specific regularisations like DropLasso may be better adapted than standard regularisations to infer molecular signatures from scRNA-seq data.

\section{Methods}

\subsection{Setting and notations}
We consider the supervised machine learning setting, where we observe a series of $n$ pairs of the form $(x_i,y_i)_{i=1,\ldots,n}$. For each $i\in[1,n]$,  $x_i \in\RR^d$ represents the gene expression levels for $d$ genes measured in the $i$-th cell by scRNA-seq, and $y_i \in \RR$ or $\{-1,1\}$ is a label to represent a discrete category or a real number associated to the $i$-th cell, e.g., a phenotype of interest such as normal vs tumour cell, or an index of progression in the cell cycle. For $i\in[1,n]$ and $j\in[1,d]$, we denote by $x_{i,j} \in \RR$ the expression level of gene $j$ in cell $i$. From this training set of $n$ annotated cells, the goal of supervised learning is to estimate a function to predict the label of any new, unseen cell from its transcriptomic profile. We restrict ourselves to linear models $f_w : \RR^d \rightarrow \RR$, for any $w\in\RR^d$, of the form
$$
\forall u\in\RR^d\,,\quad f_w(u) = \sum_{i=1}^d w_i u_i\,.
$$
To estimate a model on the training set, a popular approach is to follow a penalised maximum likelihood or empirical risk minimisation principle and to solve an objective function of the form
\begin{equation}\label{eq:main}
\min_{w \in \mathbb{R}^d}    \left\{  \frac{1}{n} \sum_{i=1}^{n} L(w, x_{i},y_{i})  + \lambda \Omega(w) \right\} \,,
\end{equation}
where $L(w, x_{i},y_{i})$ is a loss function to assess how well $f_w$ predicts $y_i$ from $x_i$, $\Omega$ is an (optional) penalty to control overfitting in high dimensions, and $\lambda>0$ is a regularisation parameter to control the balance between under- and overfitting. Examples of classical loss functions include the square loss:
 $$
 L_{\text{square}}(w,x_{i},y_{i}) =   \left(y_{i} - \sum_{j=1}^{d} w_{j}  x_{i,j}\right)  ^2  \,,
 $$
 and the logistic loss:
 $$
 L_{\text{logistic}}(w,x_{i},y_{i}) = \log  \left(1+ \exp(- y_{i}  \sum_{j=1}^{d} w_{j}  x_{i,j}) \right) \,,
 $$
 which are popular losses when $y_i$ is respectively a continuous ($y_i\in\RR$) or discrete ($y_i\in\{-1,1\}$) label. As for the regularisation term $\Omega(w)$ in (\ref{eq:main}), popular choices include the ridge penalty \citep{Hoerl1970Ridge}:
 $$
 \Omega_{\text{ridge}}(w) = \left \| w\right \|_{2}^{2} = \sum_{i=1}^d w_i^2\,,
 $$
and the lasso penalty \citep{Tibshirani1996Regression}:
$$
 \Omega_{\text{lasso}}(w) = \left \| w\right \|_{1} = \sum_{i=1}^d | w_i|\,.
 $$
 The properties, advantages and drawbacks of ridge and lasso penalties have been theoretically studied under different assumptions and regimes. The lasso penalty additionally allows  feature selection by producing sparse solutions, i.e., vectors $w$ with many zeros; this is useful to in many bioinformatics applications to select ``molecular signatures'', i.e., predictive models based on the expression of a limited number of genes only. It is known however that lasso can be unstable in particular when there are several highly correlated features in the data. It also cannot select more features than the number of observations and its accuracy is often dominated by that of ridge. For these reasons, another popular penalty is elastic net, which encompasses the advantages of both penalties \cite{Zou2005regularization} : 
$$
\Omega_{\text{ridge}}(w) = \alpha \left \| w\right \|_{2}^{2} + (1-\alpha) \|w\|_1 \,,
$$
where $\alpha\in[0,1]$ allows to interpolate between the lasso ($\alpha=0$) and the ridge ($\alpha=1$) penalties.

\subsection{DropLasso}
For scRNA-seq data subject to dropout noise, we propose a new model to train a sparse linear model robust to the noise by artificially augmenting the training set with new examples corrupted by dropout. Formally, given a vector $u \in\RR^d$ and a dropout mask $\delta \in \{0,1\}^d$, we consider the corrupted pattern $\delta \odot u \in\RR^d$ obtained by entry-wise multiplication $(\delta \odot u)_i = \delta_i u_i$. In order to consider all possible dropout masks, we make $\delta$ a random variable with independent entries following a Bernoulli distribution of parameter $p\in[0,1]$, i.e., $P(\delta_i = 1)=p$, and consider the following DropLasso regularisation for any $\lambda>0$, $p\in[0,1]$ and loss function $L$:
\begin{equation}\label{eq:droplasso}
  \min_{w \in \mathbb{R}^d}   \left( \frac{1}{n} \sum_{i=1}^{n} \underset{\delta_i \sim B(p)^d}{ \mathbb {E}}  L(w,\delta_i  \odot \frac{x_{i,}}{p}  , y_{i})   +  \lambda  \left \| w\right \|_{1} \right)  \,.
\end{equation}
In this equation, the expectation over the dropout mask corresponds to an average of $2^d$ terms. The division by $p$ in the term $x_i/p$ is here to ensure that, on average, the inner product between $w$ and $\delta_i  \odot \frac{x_{i,}}{p}$ is independent of $p$, because:
\begin{equation*}
\begin{split}
\underset{\delta_i \sim B(p)^d}{ \mathbb {E}} \sum_{j=1}^d w_j \left(\delta_i  \odot \frac{x_{i,}}{p}\right)_j &= \sum_{j=1}^d \underset{\delta_{i,j} \sim B(p)}{ \mathbb {E}}  w_j \delta_{i,j} \frac{x_{i,j}}{p} \\
&= \sum_{j=1}^d w_j x_{i,j}\,.
\end{split}
\end{equation*}
When $p=1$ and $\lambda>0$, the only mask with positive probability is the constant mask with all entries equal to $1$, which performs no dropout corruption. In that case, DropLasso (\ref{eq:droplasso}) therefore boils down to standard lasso. When $\lambda=0$ and $p<1$, on the other hand, DropLasso boils down to the standard dropout regularisation proposed by \citet{Srivastava2014Dropout} and studied, among others, by \citet{Wager2013Dropout,Baldi2013Understanding,Maaten2013Learning}. In general, DropLasso interpolates between lasso and dropout. For $\lambda>0$, it inherits from lasso regularisation the ability to select features associated with $\ell_1$ regularisation \citep{Bach2011Optimization}. We therefore propose DropLasso as a good candidate to select molecular signatures (thanks to the sparsity-inducing $\ell_1$ regularisation) for data corrupted with dropout noise, in particular scRNA-seq data (thanks to the dropout data augmentation).

\subsection{Algorithm}
For any convex loss function $L$ such as the square or logistic losses, DropLasso (\ref{eq:droplasso}) is a non-smooth convex optimisation problem whose global minimum can be found by generic solvers for convex programs. Due to the dropout corruption, the total number of terms in the sum in (\ref{eq:droplasso}) is $n\times 2^d$. This is usually prohibitive as soon as $d$ is more than a few, e.g., in practical applications when $d$ is easily of order $10^4$ (number of genes). Hence the objective function (\ref{eq:droplasso}) can simply not be computed exactly for a single candidate model $w$, and even less optimised by methods like gradient descent.

To solve (\ref{eq:droplasso}), we instead propose to follow a stochastic gradient approach to exploit the particular structure of the model, in particular the fact that it is fast and easy to generate a sample randomly corrupted by dropout noise. A similar approach is used for standard dropout regularisation when $L$ is differentiable \citet{Srivastava2014Dropout}, however in our case we additionally need to take care of the non-differentiable $\ell_1$ norm; this can be handled by a forward-backward algorithm which, plugged in the stochastic gradient loop, leads to the proximal stochastic gradient algorithm presented in Algorithm~\ref{algo:droplasso}. The fact that Algorithm~\ref{algo:droplasso} is correct, i.e., converges to the solution of (\ref{eq:droplasso}), follows from general results on stochastic approximations \citep{Robbins1971convergence}.

\begin{algorithm}[ht]
    \caption{Solving DropLasso}
    \label{algo:droplasso}
    \begin{algorithmic}[1] % The number tells where the line numbering should start
    \Require Training set $(x_i,y_i)_{i=1,\ldots,n}$, initialisation $w_0\in\RR^d$, learning rate $\gamma_0>0$, number of passes $n_{passes}\in\mathbb{N}$, $\lambda\geq 0$, $p\in[0,1]$
        \Procedure{DropLasso}{}
            \State $w^0 \gets w_0$
            \State $t \gets 0$
            \For{\texttt{$iter=1$ to $n_{passes}$}}
            \State $\pi \gets$ random permutation of $[1,n]$ \Comment{Shuffle training set}
            \For{\texttt{$i=1$ to $n$}} \Comment{(Mini-)batch also possible}
            \State $t \gets t+1$
            \State $  \gamma _t \gets \gamma_0 / (1+ \gamma_0  \lambda t)$
 %             \State Randomly pick a data point $x^{t}_{i,.}$  \\ \Comment{(Batch gradient also possible)}
              \State Sample  $\delta  \sim Bernoulli(p)^d$
              \State $z \gets \delta \odot x_{\pi(i)} / p $
%                \State Create $z_{i,.}^t$ such that  $z_{i,j}^t \gets   \frac{\delta_j x^{t}_{i,j}}{p} $
                \State $w^{t+1} \gets S_{\gamma_t \lambda}(w^{t}- \gamma_t \triangledown _{w} L(w^t,z,y_{\pi(i)}) )   $ \\  \Comment{$S_{\gamma_t \lambda} $is the soft-thresholding operator  } 
                \EndFor
            \EndFor  \\
	\Return($w^{t+1}$)
        \EndProcedure
    \end{algorithmic}
\end{algorithm}

\OMIT{
\begin{algorithm}
\caption{Working set algorithm}
\label{alg:working_set}
\begin{algorithmic}[1]
\REQUIRE $\bm{X} \in [0, 1]^{n \times p}, \TODO{or $\{0,1\}^{n\times p}$? } \bm{y} \in \RR^n$, $\lambda>0$
\ENSURE  $\bm{w}^*, b^*$
%\FOR {$t=1, 2, \dots$ until converged}
\TODO{$\theta$ initialisation missing}
\REPEAT:
\STATE $\Wcal = \left\{i \in \bbr{1,D}: \left|\bm Z_{i}^\top \bm \theta \right| \geq \lambda \right\}$ \Comment{Update the working set}
\STATE  $\bm w, b\leftarrow \underset{\bm w, b}{\text{argmin }} g_{\lambda} (\bm w, b, \bm Z_\Wcal)$ \Comment{Solve subproblem}
\STATE Compute $\bm \theta$ given  $\bm w$ and $b$. \TODO{How?}
\UNTIL convergence
\end{algorithmic}
\end{algorithm}
}

\subsection{DropLasso and elastic net}
As we already mentioned, DropLasso interpolates between lasso ($p=1,\lambda>0$) and dropout ($p\in[0,1]$, $\lambda=0$). On the other hand, dropout regularisation is known to be related to ridge regularisation \citep{Wager2013Dropout,Baldi2013Understanding}; in particular, for the square loss, dropout regularisation boils down to ridge regression after proper normalisation of the data, while for more general losses it can be approximated by reweighted version of ridge regression. Here we show that DropLasso largely inherits these properties, and in a sense is to elastic net what dropout is to ridge.
\OMIT{It is interesting to see that DropLasso bridges the Lasso and Dropout penalties as elastic net bridges Ridge and Lasso penalties. In fact, for $ \lambda=0$  DropLasso is equivalent to the dropout penalty for the case of a linear model and for $p=1$ it is equivalent to the Lasso penalty. We will see that the similarity with elastic net is stronger especially in the case of least squares. 
It is interesting to notice which properties of Lasso will be kept for DropLasso and which properties will be changed. A very special property of the Lasso estimate is its ability to produce sparse coefficients and thus to perform an embedded feature selection. 

\begin{prop}{As for Lasso, the DropLasso induces sparsity in the estimate (which depends on the tuning parameters $\lambda$ and $p$). }
\end{prop}
\paragraph{\textbf{Proof:}} {We know that DropLasso solves problem \textbf{(4)}. Setting:
 $$\widetilde{L(w,x,y)}= \underset{\delta\sim B(p)^d}{ \mathbb {E}}  L(w,\mathbf{\delta}  \odot \frac{x}{p} ,y) $$ 
Convexity of $L$ induces that of $\widetilde {L}$ , and the sparsity property of DropLasso is induced by that of Lasso for convex loss (See \cite{Bach2011Optimization} for a general analysis of Lasso estimate sparsity)  $\square $ }
}

Let us start with the square loss. In that case we have the following:
\begin{prop}{For the square loss, DropLasso is equivalent to an elastic net regression if the data are normalised so that all features have the same norm. If data are not normalised, DropLasso is equivalent to an elastic net regression with a weighted ridge penalty.}
\end{prop}

\paragraph{\textbf{Proof:}} {By developing the error function and marginalising over the Bernoulli variables we get : 

\begin{equation*} 
\begin{split}
 \frac{1}{n}  &\sum_{i=1}^{n}   \underset{\delta_{i} \sim B(p)^d}{\mathbb {E}}  L(w,{\delta_i}  \odot \frac{x_{i,}}{p}  , y_{i})   +   \lambda   \left \| w\right \|_{1}  \\
=& \frac{1}{n}  \sum_{i=1}^{n}   \underset{\delta_{i} \sim B(p)^d}{\mathbb {E}}  \left(y_{i} - \sum_{j=1}^{d} w_{j}  \delta_{i,j}  \frac{x_{i,j}}{p} \right) ^2   +   \lambda   \left \| w\right \|_{1}  
\\
%=&  \frac{1}{n}  \sum_{i=1}^{n}  \underset{\delta_{i,j} \sim B(p)}{ \mathbb {E}}  \left[ \left(  \sum_{j=1}^{d} w_{j}  \delta_{i,j}  \frac{x_{i,j}}{p} \right) ^2 - 2y_{i} \sum_{j=1}^{d} w_{j} \delta_{i,j} \frac{x_{i,j}}{p} +  y_{i} ^2 \right]  \\ &+  \lambda \left \| w\right \|_{1} \\%
 =& \frac{1}{n}   \sum_{i=1}^{n}  \left(y_{i} -\sum_{j=1}^{d} w_{j} x_{i,j} \right) ^2  + \sum_{i=1}^n \sum_{j=1}^d w_{j}^2  x_{i,j}^2   \text{Var} \left (\frac{\delta_{i,j}}{p}  \right )  +  \lambda  \left \| w\right \|_{1}    \\
 =&  \frac{1}{n} \sum_{i=1}^{n}   L(w,x_{i},y_{i}) +  \frac{1-p}{p}  \sum_{j=1}^{d}  \left \|  x_{.,j} \right \|_2^{2} w_{j}^2  +  \lambda  \left \| w\right \|_{1}.   \quad\qed
\end{split}
\end{equation*} 
 }

In the case of the logistic loss, we can also adapt a result of \citet{Wager2013Dropout} which relates dropout to an adaptive version of ridge regression:
\begin{prop}{:}
For the logistic loss, DropLasso can be approximated when the dropout probability $p$ is close to 1 by an adaptive version of elastic net that automatically scales the data but also that encourages more confident predictions. 
\end{prop}

\paragraph{\textbf{Proof:}} {Writing the Taylor expansion for the logistic loss up to the second order when the dropout is small ($p$ close to $1$), we get:
 \begin{equation*}
 \begin{aligned}
   L(w,\mathbf{\delta_i}  \odot \frac{x_{i,}}{p}  , y_{i})   & \simeq  L(w,x_{i,}  , y_{i})  \\ 
   &+ \sum_{j=1}^{d} \frac{\partial L(w,x_{i,},y)}{\partial x_{i,j}} \left ( \frac{\delta_{i,j}}{p} - 1\right )  x_{i,j}    \\ 
   &+  \frac{1}{2} \sum_{j=1}^{d}  \frac{\partial^2 L(w,x_{i,},y)}{\partial^2 x_{i,j}} \left ( \frac{\delta_{i,j}}{p} - 1 \right )^2   x_{i,j} ^2\,.
  \end{aligned}
  \end{equation*}
Taking the expectation with respect to $\delta_i$, the first order term cancels out since 
$$
\underset{\delta_i \sim B(p)^d}{ \mathbb {E}} \left [{\delta_i} \odot \frac{x_{i,}}{p} \right ] = x_{i}\,. 
$$
We then get:
 \begin{equation*}
 \begin{aligned}
 &    \frac{1}{n} \sum_{i=1}^{n} \underset{\delta_i \sim B(p)^d}{ \mathbb {E}}  L(w,{\delta_i}  \odot \frac{x_{i,}}{p}  , y_{i})   +  \lambda  \left \| w\right \|_{1}    \\
& \simeq   \sum_{i=1}^{n}  L(w,x_{i,} ,y_{i})   + \frac{1-p}{p} \sum_{j=1}^{d} \alpha_{j} w_{j} ^2 +  \lambda \left \| w\right \|_{1}  \,,
\end{aligned}
\end{equation*}
where, for the logistic loss,
 \begin{equation*}
\alpha_{j}=  %\sum_{i=1}^{n}  \frac{\partial^2 L(w,x_{i,},y)}{\partial^2 w^{\top} x_{i,j}} . x_{i,j} = %  
   \sum_{i=1}^{n}  P(Y=1 \mid X=x_i , w) P(Y=0 \mid X=x_i , w)   x^2_{i,j}  \,.\quad\qed
   \end{equation*}
  In words, this shows that the dropout penalty can be approximated by a weighted data-dependent version of ridge regression, where the ridge penalty is controlled both by the size of the features $x_{i,j}^2$, but also by the fact that the prediction for each sample is confident or not. 
   }

\section{Results}

\subsection{Simulation results}

We first investigate the performance of DropLasso on simulated data, and compare it to standard dropout and elastic net regularisation. We design a toy simulation to illustrate in particular how corruption by dropout noise impacts the performances of the different methods. The simulation goes as follow : 
\begin{itemize}
\item We set the dimension to $d=100$.
\item Each sample is a random vector $z\in\mathbf{N}^d$ with entries following a Poisson distribution with parameter 1. We introduce correlations between entries by first sampling a Gaussian copula with covariance $\Sigma_d = \mathbf{I}_d + \mathbf{1}_{d}^{\top} \mathbf{1}_d$, then transforming each entry in $[0,1]$ into an integer using the Poisson quantile function.
\item The ``true'' model is a logistic model with sparse weight vector $w\in\RR^d$ satisfying $w_i=0.05$ for $i=1,\ldots,10$ and $w_i=0$ for $i=1,\ldots,d.$
\item Using $w$ as the true underlying model and $z$ as the true observations, we simulate a label $y \sim \text{Bernoulli}( 1/(1+\exp(-\sum_{j=1}^{d} w_{j} z_{j} )) )$  
\item {We introduce corruption by dropout events by multiplying entry-wise $z$ with an i.i.d Bernoulli variables $\delta$ with probability $q$.}
\end{itemize}
We simulate $n=100$ $(z,x,y)$ samples to train different models, evaluate their performance on $m=400$ independent samples, and repeat the whole process $10$ times. Each method estimates a model using the $(x,y)$ pairs in the training set only, i.e., does only see the corrupted samples. Elastic net and DropLasso both have two parameters. In order to make a fair comparison, we fixed the $\alpha$ parameter of elastic net to $\alpha=0.5$, and the dropout probability in DropLasso to $p=0.5$. In both cases, we vary the remaining $\lambda$ parameter over a large grid of $100$ values, estimate the classification performance on the test set in terms of area under the receiving operator curve (AUC), and report the best average AUC over the grid. 
%We perform 5-folds cross-validations 10 times (2 repeats) by training on 100 labeled training examples and testing on the remaining 400, to mimic the setting of real dataset. 
\begin{table}[h]
\centering
%\processtable{Average test AUC of different methods on simulations with different amount of dropout noise. The $*$ indicates that the performance of DropLasso is significantly higher than that of elastic net ($P<0.05$) \label{Tab:01}}{
\caption{Average test AUC of different methods on simulations with different amount of dropout noise. The $*$ indicates that the performance of DropLasso is significantly higher than that of elastic net ($P<0.05$) \label{Tab:01}}{
\begin{tabular}{lllll}
\cellcolor[HTML]{C0C0C0}{\color[HTML]{333333} \textit{Method  / noise}} & \textit{no noise} & \textit{q=0.8} & \textit{q=0.6} & \textit{q=0.4} \\ \hline
\textit{Elastic net}                        & 0.641             & 0.612          & 0.557          & 0.528          \\ \hline
\textit{Dropout}                             & 0.626             & 0.613          & 0.551          & 0.525          \\ \hline
\textit{DropLasso}                         & \textbf{0.642}    & \textbf{0.625} & \textbf{0.565 } & \textbf{0.542 *}
\end{tabular}}{}
\end{table}

Table \ref{Tab:01} shows the classification performance in terms of best average AUC of elastic net, dropout and DropLasso, when we vary the amount of dropout corruption in the data.
We first observe that for all methods, the performance drastically decreases when dropout noise increases, confirming the difficulty induced by dropout events to learn predictive models. 
Second, we note that, whatever the amount of noise, DropLasso outperforms dropout. This illustrates the benefit of incorporating the $\ell_1$ lasso penalty in the objective function of dropout when the true model is sparse. Third, and more importantly, we observe that DropLasso outperforms elastic net in all settings, and that the difference in performance increases when the amount of dropout noise increases. This confirms the intuition that DropLasso is more efficient that elastic net in situations where data are corrupted by dropout noise.

\begin{figure*}
\centering{
  \includegraphics[scale=0.34]{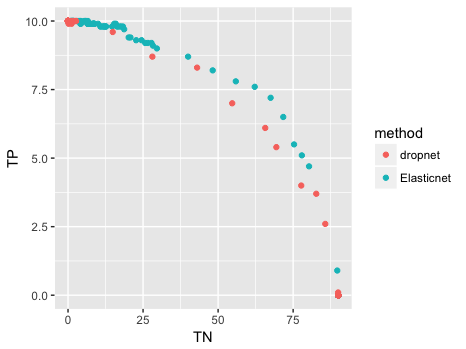}
  \includegraphics[scale=0.34]{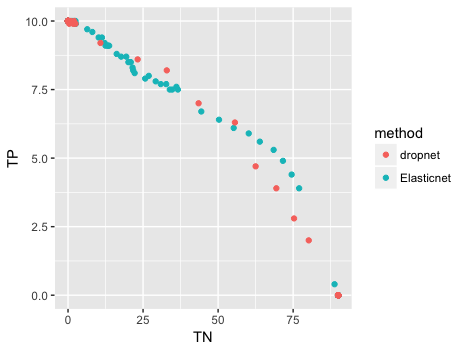}
  \includegraphics[scale=0.34]{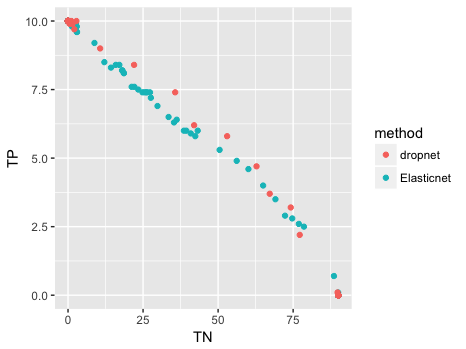}
}
   \caption{Performance on feature selection for elastic net and DropLasso on simulated data. From left to right, the amount of dropout noise in the data increases from no noise (left, $q=1$), to $q=0.8$ (centre) and $q=0.4$ (right)\label{fig:simfeat}}
\end{figure*}

Besides classification accuracy, it is also of interest to investigate to what extent the different methods select the correct variables, which are known in our simulations. For each $\lambda$ value in the grid, we compute the number of true and false positives among the features selected by elastic net and DropLasso, and plot these values averaged over the 10 repeats in Figure~\ref{fig:simfeat}.
Interestingly, we observe that elastic net seems to outperform DropLasso when there is no noise in the observed data ($q=1$), but seems to lose its ability to recover the correct features as the amount of dropout noise increases quicker than DropLasso, and in the high noise regime ($q=0.4$) DropLasso eventually outperforms elastic net. Although the differences are limited in this simulation setting, this illustrates again that DropLasso is more robust than elastic net in the presence of dropout noise.

\subsection{Classification on Single Cell RNA-seq}

%\subsubsection{Data and setting}
We now turn on to real scRNA-seq data. To evaluate the performance of methods for supervised classification, we collected $7$ publicly available scRNA-seq datasets amenable to this setting, as summarised in Table~\ref{tab-data}. The first 6 datasets were processed \citet{Soneson2017Bias}, and we obtained them from the \textit{conquer} website\footnote{\href{http://imlspenticton.uzh.ch:3838/conquer/}{http://imlspenticton.uzh.ch:3838/conquer/}}, a collection of consistently processed, analysis-ready and well documented publicly available scRNA-seq data sets. We used the already preprocessed length-scaled transcripts per million mapped reads \citep[see][for details about data processing]{Soneson2017Bias}. These datasets were used by \citet{Soneson2017Bias} to assess the performance of methods for gene differential analysis between classes of cells, and we follow the same splits of cells into classes for our experiments of supervised classification. The last dataset was collected from \citet{Li2017Reference} and was originally used for the analysis of transcriptional heterogeneity in colorectal tumours. Gene expression levels were quantified as fragments per kilo-base per million reads (FPKM) which also allowed to compare the classifiers after different normalisations. We used the available sample annotations to create a binary classification problem where we want to discriminate tumour from normal epithelial cells, as described in Table~\ref{tab-data}. For all datasets, we filtered out genes that were not expressed in any sample. In order to also study the performance of the different methods for a varying number of variables, we also created for each dataset 3 reduced datasets by first selecting respectively the top 100, 1,000 and 10,000 genes with highest variance among the samples, regardless of their labels. 
%We do not claim this is the best dimension reduction one could use for this data (even if it is interesting to notice that Signal to noise ration as reflected in the methods classification performance seems to be higher with less genes which indicates the presence of many noisy variables or at least the dominance of noise over signals for many variables.) 
On each of the $7\times 4=28$ resulting dataset, we compare the performance of 4 regularisation methods for logistic regression: lasso, dropout, elastic net and DropLasso. As in the simulation study, we fix the probability of dropout to $p=0.5$ for dropout and DropLasso regularisation. For elastic net, we fix $\alpha=0.5$ to balance the $\ell_1$ and $\ell_2$ norms. Finally, for lasso, elastic net and DropLasso, we run the models with 100 values for $\lambda$, regularly spaced after log transform between $\lambda_{\text{min}}=10^{-5}$ and $\lambda_{\text{max}}=10^5$, on 20\% of the data chosen in such way that labels are balanced, and evaluate the performance of the resulting models on the 80\% remaining data. We report in Table~\ref{tab-data} the maximal test AUC, averaged over the 5 repeats, taken over the grid of $\lambda$.
 \begin{table}[h]
 \centering
% \processtable{Experimental datasets \label{tab-data}}{
 \caption{Experimental datasets \label{tab-data}}{
\scalebox{0.7}{
\begin{tabular}{@{}llllll@{}}
\toprule
\textbf{Dataset}  & \textbf{Classification Task}                                                                                                                                     & \textbf{Variables} & \textbf{Samples} & \textbf{Organism} \\ \midrule
EMTAB2805        & G1 vs G2M                                                                                                                                                        & 20 614             & 96 ; 96          & mouse                 \\
GSE74596          & NKT0 vs NKT17                                                                                                                                                    & 14 172             & 44 ; 45          & mouse                \\
GSE45719          & 16-cell stage blastomere vs midblastocyst                                                                                                                    & 21 605             & 50 ; 60          & mouse                \\
GSE63818-GPL16791 & Primordial Germ Cells vs Somatic Cells  								& 30 022             & 26 ; 40          & human             &                    \\
GSE48968-GPL13112 & BMDC:  1h LPS vs 4h LPS Stimulation                                                                                                         & 17 947             & 95 ; 96          & mouse                       \\
GSE60749-GPL13112 & Culture condition: 2i+LIF vs Serum+LIF                                              & 39 351             & 90 ; 94          & mouse                      \\
GSE81861          & Epithelial cells: Tumour (colorectal) vs Normal                                                                                                                  & 36 400             & 160 ; 160        & human              \\ \bottomrule
\end{tabular}
}}{}
\end{table}

On each of the $7\times 4=28$ resulting dataset, we compare the performance of 4 regularisation methods for logistic regression: lasso, dropout, elastic net and DropLasso. As in the simulation study, we fix the probability of dropout to $p=0.5$ for dropout and DropLasso regularisation. For elastic net, we fix $\alpha=0.5$ to balance the $\ell_1$ and $\ell_2$ norms. Finally, for lasso, elastic net and DropLasso, we run the models with 100 values for $\lambda$, regularly spaced after log transform between $\lambda_{\text{min}}=10^{-5}$ and $\lambda_{\text{max}}=10^5$, on 20\% of the data chosen in such way that labels are balanced, and evaluate the performance of the resulting models on the 80\% remaining data. We report in Table~\ref{tab-data} the maximal test AUC, averaged over the 5 repeats, taken over the grid of $\lambda$. 
\begin{table}[h]
% \processtable{Best average test AUC score for lasso, dropout, elastic net and DropLasso regularised logistic regression on 7 datasets and their 3 reduced datasets. $*$ and $**$ indicate when the performance of DropLasso is significantly better than that of elastic net ($P<0.05$ and $P<0.01$, respectively) \label{tab-auc_exp}}{
\centering
 \caption{Best average test AUC score for lasso, dropout, elastic net and DropLasso regularised logistic regression on 7 datasets and their 3 reduced datasets. $*$ and $**$ indicate when the performance of DropLasso is significantly better than that of elastic net ($P<0.05$ and $P<0.01$, respectively) \label{tab-auc_exp}}{
\scalebox{0.8}{
\begin{tabular}{@{}lllllll@{}}
\toprule
\textbf{Dataset}           & \textbf{Number of variables} & \multicolumn{1}{c}{\textbf{LASSO}} & \textbf{Dropout} & \textbf{Elastic net} & \textbf{DropLasso}  \\ \midrule
EMTAB2805       & 100                          & 0.95                               & 0.94             & \textbf{0.966}      & 0.964                                \\
                           & 1 000                        & 0.956                              & 0.989            & 0.980               & \textbf{0.990 *}                        \\
                           & 10 000                       & 0.764                              & 0.961            & 0.817               & \textbf{0.961 *}                             \\
                           & All (20 614)                 & 0.72                               & 0.928            & 0.796               & \textbf{0.946 **}                         \\ \midrule
GSE74596         & 100                          & 0.997                              & 0.996            & 0.994               & \textbf{0.998}                    	        \\
                           & 1 000                        & 0.988                              & 0.997            & 0.994               & \textbf{0.999}                          \\
                           & 10 000                       & 0.769                              & 0.960            & 0.909               & \textbf{0.990*}                            \\
                           & All (14 172)                 & 0.844                              & 0.915            & 0.943               & \textbf{0.966}                            \\ \midrule
GSE45719         & 100                          & 0.999                              & 0.990            & 0.999               & \textbf{0.999}                       \\
                           & 1 000                        &       0.997                            & 0.999            & 0.999               & \textbf{1}                              \\
                           & 10 000                       & 0.995                              & 0.998            & 0.998               & \textbf{1 *}                            \\
                           & All                          & 0.990                              & 0.999            & 0.999               & \textbf{1}                              \\ \midrule
GSE63818-GPL16791 & 100                          & 0.94                               & 0.977            & 0.984               & \textbf{0.998 *}                        \\
                           & 1 000                        & 0.945                              & 0.998            & 0.985               & \textbf{1 *}                                \\
                           & 10 000                       & 0.951                              & 0.995            & 0.987               & \textbf{0.998 *}                         \\
                           & All                          & 0.932                              & 0.970            & 0.976               & \textbf{0.989}                         \\ \midrule
GSE48968-GPL13112          & 100                          & 0.995                              & 0.992            & 0.996               & \textbf{0.997}                            \\
                           & 1 000                        & 0.962                              & 0.992            & 0.996               & \textbf{0.997}                            \\
                           & 10 000                       & 0.939                              & 0.97             & 0.978               & \textbf{0.992 *}                           \\
                           & All                          & 0.948                              & 0.962            & 0.96                & \textbf{0.987 *}                          \\ \midrule
GSE60749-GPL13112          & 100                          & \textbf{1}                         & \textbf{1}       & \textbf{1}          & \textbf{1}                        \\
                           & 1 000                        & \textbf{1}                         & 0.998            & \textbf{1}          & \textbf{1}                         \\
                           & 10 000                       & \textbf{1}                         & 0.999            & \textbf{1}          & \textbf{1}                         \\
                           & All                          & \textbf{1}                         & 0.997            & \textbf{1}          & \textbf{1}                                  \\ \midrule
GSE81861                   & 100                          & 0.852                              & 0.808            & 0.854               & \textbf{0.887 *}                             \\
                           & 1 000                        & 0.89                               & 0.915            & 0.898               & \textbf{0.933 *}                        \\
                           & 10 000                       & 0.827                              & 0.9              & 0.881               & \textbf{0.927 **}                          \\
                           & All                          & 0.8                                & 0.851            & 0.84                & \textbf{0.9 **}                          \\ \bottomrule
\end{tabular}
}}{}
\end{table}

\begin{figure}[h]
  \centering{
 \includegraphics[width=0.4\textwidth]{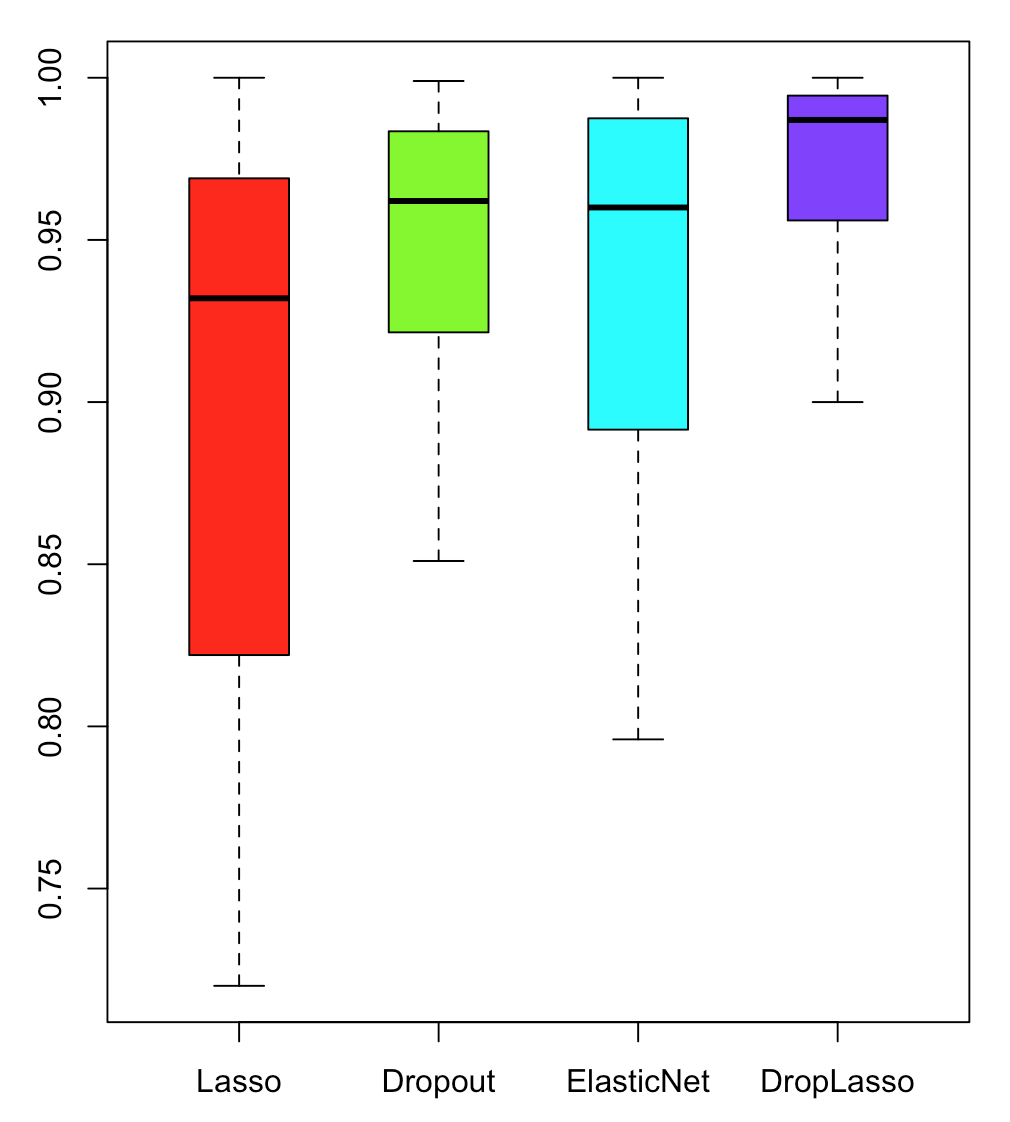}
\caption{Average CV-AUC for all 7 datasets}\label{fig:boxplot}}
\end{figure}

The first observation is that the performances reached by all methods on all datasets are generally very high, and can reach an AUC above 0.9 on each of the 7 datasets. This suggests that the labels chosen in these datasets are sufficiently different in terms of transcriptomic profiles that they can be easily recognised most of the time. We still notice some differences in performance between datasets, with GSE60749-GPL13112 being the easiest one while EMTAB2805 is the most challenging, for all methods. \citet{Soneson2017Bias} also noticed a difference in signal-to-noise ratio between these datasets, in the context of gene differential analysis. Second, we observe that the best performance is obtained by DropLasso on 27 out of the 28 datasets. The difference between the 4 regularisers is visualised in the box plots on Figure~\ref{fig:boxplot}, which summarise the AUC values over the 7 experiments with all genes for each method. In 14 of the 28 experiments, DropLasso significantly outperforms elastic net at significance level $P<0.05$, and in 3 of these cases the significance level is $P<0.01$.
This confirms that on real scRNA-seq data, DropLasso also brings a consistent benefit over dropout of elastic net regularisation. Finally, regarding the impact of the number of features selected, we observe that in most datasets and for most methods the best performance is reached for 100 or 1,000 genes. Remembering that genes are only selected based on their variance, independently of any class label information, this suggests that all method suffer in the high-dimensional regime and that a simple pre-filtering of genes can help by reducing the dimension of the problem. This also indirectly confirms the importance of regularisation in high dimension, and the relevance of developing adequate regularisers incorporating prior knowledge about the data and their noise.

%\subsubsection{Biological significance of the selected features:}
To conclude this section, we now compare the lists of genes in the molecular signatures estimated by elastic net and DropLasso regularisation. We illustrate this comparison on the first dataset, EMTAB2805, where the goal is to discriminate mouse cells at the G1 from the G2M cell cycle stages. To this end, we retrain the different methods with the tuning parameter $\lambda$ corresponding to the best accuracy on the 1000-filtered datasets with all the samples, and then we perform a bioinformatic analysis of Gene Ontology annotations using DAVID \citep{Huang2009Bioinformatics} on the subset of genes with non-zero coefficients for each method.

For this dataset and the best tuning parameters, DropLasso selects 186 variables while elastic net only selects 48 variables. The analysis of the selected genes shows enrichment in the functional terms ``cell division'', ``cell cycle'' and ``mitosis'' for both methods. For DropLasso, 21 genes are related to the functional term ``cell division''. Among those, one can find CDC28, CDC23 and CDca8 which were not selected by elastic net, that only selected 10 genes related to cell division. Similarly 30 genes that have previously been annotated with cellular processes consistent with the ``cell cycle'' term are selected in DropLasso versus 9 in elastic net, and 22 genes related to ``mitosis'' versus 10 in elastic net. Therefore, DropLasso could potentially allow for the discovery of more functionally related genes in its signature than elastic net.

\section{Discussion}

ScRNA-seq is changing the way we study cellular heterogeneity and investigate a number of biological process such as differentiation or tumourigenesis. Yet, as the throughput of scRNA-seq technologies increases and allows to process more and more cells simultaneously, it is likely that the amount of information captured in each individual cell will remain limited in the future and that dropout noise will continue to affect scRNA-seq (and other single-cell technologies).

Several techniques have been proposed to handle dropout noise in the context of data normalisation or gene differential expression analysis, and shown to outperform standard techniques widely used for bulk RNA-seq data analysis. In this paper we investigate a new setting which, we believe, will play an important role in the future: supervised classification of cell populations into pre-specified classes, and selection of molecular signatures for that purpose. Molecular signatures for the classification of tissues from bulk RNA-seq data has already had a tremendous impact in cancer research, and as more and more cell types are investigated and discovered with scRNA-seq it is likely that specific molecular signatures will be useful in the future to automatically sort cells into their classes.

DropLasso, the new technique we propose, borrows the recent idea of dropout regularisation from machine learning, and extends it to allow feature selection. While a parallel between dropout regularisation and (data-dependent) ridge regression has already been shown by \citet{Wager2013Dropout} and \citet{Baldi2013Understanding}, it is reassuring that we are able to extend this parallel to DropLasso and elastic net regularisation.

More interesting is the fact that, on both simulated and real data, we obtained promising results with DropLasso. They suggest that, again, specific models tailored to the data and noise give an edge over generic models developed under different assumptions.

The intuition behind why dropout (and DropLasso) perform well on scRNA-seq data, however, remains a bit unclear. Our main motivation to use them in this context was to see them as data augmentation techniques, where training data are corrupted according to the noise we assume in the data. While we believe this is fundamentally the reason why we obtained promising results, alternative explanations for the success of dropout have been proposed, and may also play a role in the context of scRNA-seq. They include for example the interpretation of dropout as a regulariser similar to a data-dependent weighted version of ridge regularisation, which works well in the presence of rare but important features \citep{Wager2013Dropout}; it would be interesting to clarify if the regularisation induced by DropLasso on scRNA-seq data exploits some fundamental property of these data, and may be replaced by a more direct approach to model this.

Finally, this first study of dropout and DropLasso regularisation on biological data paves the way to many future direction. For example, it is known that the probability of dropout in scRNA-seq data depends on the gene expression level \citep{Kharchenko2014Bayesian,Risso2018ZINB}. It would therefore be interesting to study both theoretically and empirically if a dropout regularisation following a similar pattern may be useful. Second,  instead of independently perturbing the different features one may create a correlation between the dropout events in different genes. Creating a correlation may be a way to create new regularisation by generating a \emph{structured} dropout noise. It may for example be possible to derive a correlation structure for dropout noise from prior knowledge about gene annotations or gene networks in order to enforce a structure in the molecular signature, just like structured ridge and lasso penalties have been used to promote structure molecular signatures with bulk transcriptomes \citep{Rapaport2007Classification,Jacob2009Group}.

%%%%%%%%%%%%%%%%%%%%%%%%%%%%%%%%%%%%%%%%%%%%%%%%%%%%%%%%%%%%%%%%%%%%%%%%%%%%%%%%%%%%%
%
%     please remove the " % " symbol from \centerline{\includegraphics{fig01.eps}}
%     as it may ignore the figures.
%
%%%%%%%%%%%%%%%%%%%%%%%%%%%%%%%%%%%%%%%%%%%%%%%%%%%%%%%%%%%%%%%%%%%%%%%%%%%%%%%%%%%%%%

%\section{Conclusion}

\section*{Funding}

This work has been supported by the European Research Council (grant ERC-SMAC-280032).\vspace*{-12pt}

%\bibliographystyle{natbib}
%\bibliographystyle{achemnat}
%\bibliographystyle{plainnat}
%\bibliographystyle{abbrv}
%\bibliographystyle{bioinformatics}
%
%\bibliographystyle{plain}
%
%\bibliography{Document}

\begin{thebibliography}{}

\bibitem[Bach {\em et~al.}(2011)Bach, Jenatton, Mairal, and
  Obozinski]{Bach2011Optimization}
Bach, F., Jenatton, R., Mairal, J., and Obozinski, G. (2011).
\newblock Optimization with sparsity-inducing penalties.
\newblock {\em Foundations and Trends{\textregistered} in Machine Learning\/},
  {\bf 4}(1), 1--106.

\bibitem[Bacher and Kendziorski(2016)Bacher and Kendziorski]{Bacher2016Design}
Bacher, R. and Kendziorski, C. (2016).
\newblock {Design and computational analysis of single-cell RNA-sequencing
  experiments}.
\newblock {\em Genome Biology\/}, {\bf 17}(63).

\bibitem[Baldi and Sadowski(2013)Baldi and Sadowski]{Baldi2013Understanding}
Baldi, P. and Sadowski, P.~J. (2013).
\newblock Understanding dropout.
\newblock In C.~J.~C. Burges, L.~Bottou, M.~Welling, Z.~Ghahramani, and K.~Q.
  Weinberger, editors, {\em Adv. Neural. Inform. Process Syst.}, pages
  2814--2822. Curran Associates, Inc.

\bibitem[Breiman(1996)Breiman]{Breiman1996Bagging}
Breiman, L. (1996).
\newblock Bagging predictors.
\newblock {\em Mach. Learn.}, {\bf 24}(2), 123--140.

\bibitem[Breiman(2001)Breiman]{Breiman2001Random}
Breiman, L. (2001).
\newblock Random forests.
\newblock {\em Mach. Learn.}, {\bf 45}(1), 5--32.

\bibitem[Deng {\em et~al.}(2014)Deng, Ramsk{\"{o}}ld, Reinius, and
  Sandberg]{Deng2014Single}
Deng, Q., Ramsk{\"{o}}ld, D., Reinius, B., and Sandberg, R. (2014).
\newblock {Single-cell RNA-seq reveals dynamic, random monoallelic gene
  expression in mammalian cells}.
\newblock {\em Science\/}, {\bf 343}(6167), 193--6.

\bibitem[Haury and Vert(2010)Haury and Vert]{Haury2010stability}
Haury, A.-C. and Vert, J.-P. (2010).
\newblock On the stability and interpretability of prognosis signatures in
  breast cancer.
\newblock In {\em Proceedings of the Fourth International Workshop on Machine
  Learning in Systems Biology (MLSB10)\/}.
\newblock To appear.

\bibitem[Hoerl and Kennard(1970)Hoerl and Kennard]{Hoerl1970Ridge}
Hoerl, A.~E. and Kennard, R.~W. (1970).
\newblock Ridge regression~: biased estimation for nonorthogonal problems.
\newblock {\em Technometrics\/}, {\bf 12}(1), 55--67.

\bibitem[Huang {\em et~al.}(2009)Huang, Sherman, and
  Lempicki]{Huang2009Bioinformatics}
Huang, D.~W., Sherman, B.~T., and Lempicki, R.~A. (2009).
\newblock Bioinformatics enrichment tools: paths toward the comprehensive
  functional analysis of large gene lists.
\newblock {\em Nucl. Acids Res.}, {\bf 37}, 1--13.

\bibitem[Jacob {\em et~al.}(2009)Jacob, Obozinski, and Vert]{Jacob2009Group}
Jacob, L., Obozinski, G., and Vert, J.-P. (2009).
\newblock Group lasso with overlap and graph lasso.
\newblock In {\em ICML '09: Proceedings of the 26th Annual International
  Conference on Machine Learning\/}, pages 433--440, New York, NY, USA. ACM.

\bibitem[Kharchenko {\em et~al.}(2014)Kharchenko, Silberstein, and
  Scadden]{Kharchenko2014Bayesian}
Kharchenko, P.~V., Silberstein, L., and Scadden, D.~T. (2014).
\newblock Bayesian approach to single-cell differential expression analysis.
\newblock {\em Nat. Methods\/}, {\bf 11}(7), 740--742.

\bibitem[Kolodziejczyk {\em et~al.}(2015)Kolodziejczyk, Kim, Svensson, Marioni,
  and Teichmann]{Kolodziejczyk2015technology}
Kolodziejczyk, A.~A., Kim, J.~K., Svensson, V., Marioni, J.~C., and Teichmann,
  S.~A. (2015).
\newblock {The technology and biology of single-cell RNA sequencing}.
\newblock {\em Molecular Cell\/}, {\bf 58}(4), 610--620.

\bibitem[Krizhevsky {\em et~al.}(2012)Krizhevsky, Sutskever, and
  Hinton]{Krizhevsky2012ImageNet}
Krizhevsky, A., Sutskever, I., and Hinton, G.~E. (2012).
\newblock {ImageNet} classification with deep convolutional neural networks.
\newblock In F.~Pereira, C.~Burges, L.~Bottou, and K.~Weinberger, editors, {\em
  Adv. Neural. Inform. Process Syst.}, volume~25, pages 1097--1105. Curran
  Associates, Inc.

\bibitem[Li {\em et~al.}(2017)Li, Courtois, Sengupta, Tan, Chen, Goh, Kong,
  Chua, Hon, and Tan]{Li2017Reference}
Li, H., Courtois, E.~T., Sengupta, D., Tan, Y., Chen, K.~H., Goh, J. J.~L.,
  Kong, S.~L., Chua, C., Hon, L.~K., and Tan, W.~S. (2017).
\newblock Reference component analysis of single-cell transcriptomes elucidates
  cellular heterogeneity in human colorectal tumors.
\newblock {\em Nat. Genet.}, {\bf 49}(5), 708.

\bibitem[Macosko {\em et~al.}(2015)Macosko, Basu, Satija, Nemesh, Shekhar,
  Goldman, Tirosh, Bialas, Kamitaki, Martersteck, Trombetta, Weitz, Sanes,
  Shalek, Regev, and McCarroll]{Macosko2015Highly}
Macosko, E.~Z., Basu, A., Satija, R., Nemesh, J., Shekhar, K., Goldman, M.,
  Tirosh, I., Bialas, A.~R., Kamitaki, N., Martersteck, E.~M., Trombetta,
  J.~J., Weitz, D.~A., Sanes, J.~R., Shalek, A.~K., Regev, A., and McCarroll,
  S.~A. (2015).
\newblock {Highly Parallel Genome-wide Expression Profiling of Individual Cells
  Using Nanoliter Droplets}.
\newblock {\em Cell\/}, {\bf 161}(5), 1202--1214.

\bibitem[Ozsolak and Milos(2011)Ozsolak and Milos]{Ozsolak2011RNA}
Ozsolak, F. and Milos, P.~M. (2011).
\newblock {RNA} sequencing: advances, challenges and opportunities.
\newblock {\em Nat. Rev. Genet.}, {\bf 12}, 87--98.

\bibitem[Patel {\em et~al.}(2014)Patel, Tirosh, Trombetta, Shalek, Gillespie,
  Wakimoto, Cahill, Nahed, Curry, Martuza, Louis, Rozenblatt-Rosen, Suv\`a,
  Regev, and Bernstein]{Patel2014Single}
Patel, A.~P., Tirosh, I., Trombetta, J.~J., Shalek, A.~K., Gillespie, S.~M.,
  Wakimoto, H., Cahill, D.~P., Nahed, B.~V., Curry, W.~T., Martuza, R.~L.,
  Louis, D.~N., Rozenblatt-Rosen, O., Suv\`a, M.~L., Regev, A., and Bernstein,
  B.~E. (2014).
\newblock Single-cell {RNA}-seq highlights intratumoral heterogeneity in
  primary glioblastoma.
\newblock {\em Science\/}, {\bf 344}(6190), 1396--1401.

\bibitem[Perou {\em et~al.}(2000)Perou, S{\o}rlie, Eisen, van~de Rijn, Jeffrey,
  Rees, Pollack, Ross, Johnsen, Akslen, Fluge, Pergamenschikov, Williams, Zhu,
  L{\o}nning, B{\o}rresen-Dale, Brown, and Botstein]{Perou2000Molecular}
Perou, C.~M., S{\o}rlie, T., Eisen, M.~B., van~de Rijn, M., Jeffrey, S.~S.,
  Rees, C.~A., Pollack, J.~R., Ross, D.~T., Johnsen, H., Akslen, L.~A., Fluge,
  O., Pergamenschikov, A., Williams, C., Zhu, S.~X., L{\o}nning, P.~E.,
  B{\o}rresen-Dale, A.~L., Brown, P.~O., and Botstein, D. (2000).
\newblock Molecular portraits of human breast tumours.
\newblock {\em Nature\/}, {\bf 406}(6797), 747--752.

\bibitem[Pierson and Yau(2015)Pierson and Yau]{Pierson2015Dimensionality}
Pierson, E. and Yau, C. (2015).
\newblock Dimensionality reduction for zero-inflated single cell gene
  expression analysis.
\newblock {\em Genome Biol.}, {\bf 16}(241).

\bibitem[Ramaswamy {\em et~al.}(2001)Ramaswamy, Tamayo, Rifkin, Mukherjee,
  Yeang, Angelo, Ladd, Reich, Latulippe, Mesirov, Poggio, Gerald, Loda, Lander,
  and Golub]{Ramaswamy2001Multiclass}
Ramaswamy, S., Tamayo, P., Rifkin, R., Mukherjee, S., Yeang, C., Angelo, M.,
  Ladd, C., Reich, M., Latulippe, E., Mesirov, J., Poggio, T., Gerald, W.,
  Loda, M., Lander, E., and Golub, T. (2001).
\newblock Multiclass cancer diagnosis using tumor gene expression signatures.
\newblock {\em Proc. {N}atl. {A}cad. {S}ci. {USA}\/}, {\bf 98}(26),
  15149--15154.

\bibitem[Rapaport {\em et~al.}(2007)Rapaport, Zynoviev, Dutreix, Barillot, and
  Vert]{Rapaport2007Classification}
Rapaport, F., Zynoviev, A., Dutreix, M., Barillot, E., and Vert, J.-P. (2007).
\newblock Classification of microarray data using gene networks.
\newblock {\em BMC Bioinformatics\/}, {\bf 8}, 35.

\bibitem[Risso {\em et~al.}(2018)Risso, Perraudeau, Gribkova, Dudoit, and
  Vert]{Risso2018ZINB}
Risso, D., Perraudeau, F., Gribkova, S., Dudoit, S., and Vert, J.-P. (2018).
\newblock {ZINB-WaVE: A general and flexible method for signal extraction from
  single-cell RNA-seq data}.
\newblock {\em Nature Comm.}, {\bf 9}(1), 284.

\bibitem[Robbins and Siegmund(1971)Robbins and
  Siegmund]{Robbins1971convergence}
Robbins, H. and Siegmund, D. (1971).
\newblock A convergence theorem for non negative almost supermartingales and
  some applications.
\newblock In {\em Optimizing methods in statistics\/}, pages 233--257.
  Elsevier.

\bibitem[Sch{\"o}lkopf {\em et~al.}(1996)Sch{\"o}lkopf, Burges, and
  Vapnik]{Scholkopf1996Incorporating}
Sch{\"o}lkopf, B., Burges, C., and Vapnik, V. (1996).
\newblock Incorporating invariances in support vector learning machines.
\newblock In C.~von~der Malsburg, W.~von Seelen, J.~C. Vorbr\"{u}ggen, and
  B.~Sendhoff, editors, {\em ICANN 96: Proceedings of the 1996 International
  Conference on Artificial Neural Networks\/}, pages 47--52, London, UK.
  Springer-Verlag.

\bibitem[Soneson and Robinson(2017)Soneson and Robinson]{Soneson2017Bias}
Soneson, C. and Robinson, M.~D. (2017).
\newblock Bias, robustness and scalability in differential expression analysis
  of single-cell {RNA}-seq data.
\newblock Technical Report 143289, bioRxiv.

\bibitem[S{\o}rlie {\em et~al.}(2001)S{\o}rlie, Perou, Tibshirani, Aas,
  Geisler, Johnsen, Hastie, Eisen, van~de Rijn, Jeffrey, Thorsen, Quist,
  Matese, Brown, Botstein, Eystein~L{\o}nning, and
  B{\o}rresen-Dale]{Soerlie2001Gene}
S{\o}rlie, T., Perou, C.~M., Tibshirani, R., Aas, T., Geisler, S., Johnsen, H.,
  Hastie, T., Eisen, M.~B., van~de Rijn, M., Jeffrey, S.~S., Thorsen, T.,
  Quist, H., Matese, J.~C., Brown, P.~O., Botstein, D., Eystein~L{\o}nning, P.,
  and B{\o}rresen-Dale, A.~L. (2001).
\newblock Gene expression patterns of breast carcinomas distinguish tumor
  subclasses with clinical implications.
\newblock {\em Proc. Natl. Acad. Sci. USA\/}, {\bf 98}(19), 10869--10874.

\bibitem[S{\o}rlie {\em et~al.}(2003)S{\o}rlie, Tibshirani, Parker, Hastie,
  Marron, Nobel, Deng, Johnsen, Pesich, Geisler, Demeter, Perou, Lønning,
  Brown, Børresen-Dale, and Botstein]{Sorlie2003Repeated}
S{\o}rlie, T., Tibshirani, R., Parker, J., Hastie, T., Marron, J., Nobel, A.,
  Deng, S., Johnsen, H., Pesich, R., Geisler, S., Demeter, J., Perou, C.,
  Lønning, P., Brown, P., Børresen-Dale, A., and Botstein, D. (2003).
\newblock Repeated observation of breast tumor subtypes in independent gene
  expression data sets.
\newblock {\em Proc. Natl. Acad. Sci. USA\/}, {\bf 100}(14), 8418--8423.

\bibitem[Srivastava {\em et~al.}(2014)Srivastava, Hinton, Krizhevsky,
  Sutskever, and Salakhutdinov]{Srivastava2014Dropout}
Srivastava, N., Hinton, G., Krizhevsky, A., Sutskever, I., and Salakhutdinov,
  R. (2014).
\newblock Dropout: A simple way to prevent neural networks from overfitting.
\newblock {\em J. Mach. Learn. Res.}, {\bf 15}(1), 1929--1958.

\bibitem[Tasic {\em et~al.}(2016)Tasic, Menon, Nguyen, Kim, Jarsky, Yao, Levi,
  Gray, Sorensen, Dolbeare, Bertagnolli, Goldy, Shapovalova, Parry, Lee, Smith,
  Bernard, Madisen, Sunkin, Hawrylycz, Koch, and Zeng]{Tasic2016Adult}
Tasic, B., Menon, V., Nguyen, T.~N., Kim, T.~K., Jarsky, T., Yao, Z., Levi, B.,
  Gray, L.~T., Sorensen, S.~A., Dolbeare, T., Bertagnolli, D., Goldy, J.,
  Shapovalova, N., Parry, S., Lee, C., Smith, K., Bernard, A., Madisen, L.,
  Sunkin, S.~M., Hawrylycz, M., Koch, C., and Zeng, H. (2016).
\newblock Adult mouse cortical cell taxonomy revealed by single cell
  transcriptomics.
\newblock {\em Nat. Neurosci.}, {\bf 19}(2), 335--346.

\bibitem[Tibshirani(1996)Tibshirani]{Tibshirani1996Regression}
Tibshirani, R. (1996).
\newblock Regression shrinkage and selection via the lasso.
\newblock {\em J. R. Stat. Soc. Ser. B\/}, {\bf 58}(1), 267--288.

\bibitem[van~de Vijver {\em et~al.}(2002)van~de Vijver, He, van't Veer, Dai,
  Hart, Voskuil, Schreiber, Peterse, Roberts, Marton, Parrish, Atsma,
  Witteveen, Glas, Delahaye, van~der Velde, Bartelink, Rodenhuis, Rutgers,
  Friend, and Bernards]{Vijver2002gene-expression}
van~de Vijver, M.~J., He, Y.~D., van't Veer, L.~J., Dai, H., Hart, A. A.~M.,
  Voskuil, D.~W., Schreiber, G.~J., Peterse, J.~L., Roberts, C., Marton, M.~J.,
  Parrish, M., Atsma, D., Witteveen, A., Glas, A., Delahaye, L., van~der Velde,
  T., Bartelink, H., Rodenhuis, S., Rutgers, E.~T., Friend, S.~H., and
  Bernards, R. (2002).
\newblock A gene-expression signature as a predictor of survival in breast
  cancer.
\newblock {\em N. Engl. J. Med.}, {\bf 347}(25), 1999--2009.

\bibitem[van~der Maaten {\em et~al.}(2013)van~der Maaten, Chen, Tyree, and
  Weinberger]{Maaten2013Learning}
van~der Maaten, L., Chen, M., Tyree, S., and Weinberger, K.~Q. (2013).
\newblock Learning with marginalized corrupted features.
\newblock In {\em Proceedings of the 30th International Conference on Machine
  Learning, ICML 2013, Atlanta, GA, USA, 16-21 June 2013\/}, number~28 in JMLR
  Proceedings, pages 410--418. JMLR.org.

\bibitem[Villani {\em et~al.}(2017)Villani, Satija, Reynolds, Sarkizova,
  Shekhar, Fletcher, Griesbeck, Butler, Zheng, Lazo, {\em
  et~al.}]{Villani2017Single}
Villani, A.-C., Satija, R., Reynolds, G., Sarkizova, S., Shekhar, K., Fletcher,
  J., Griesbeck, M., Butler, A., Zheng, S., Lazo, S., {\em et~al.} (2017).
\newblock Single-cell {RNA}-seq reveals new types of human blood dendritic
  cells, monocytes, and progenitors.
\newblock {\em Science\/}, {\bf 356}(6335), eaah4573.

\bibitem[Wager {\em et~al.}(2013)Wager, Wang, and Liang]{Wager2013Dropout}
Wager, S., Wang, S., and Liang, P.~S. (2013).
\newblock Dropout training as adaptive regularization.
\newblock In C.~Burges, L.~Bottou, M.~Welling, Z.~Ghahramani, and
  K.~Weinberger, editors, {\em Adv. Neural. Inform. Process Syst.}, volume~26,
  pages 351--359. Curran Associates, Inc.

\bibitem[Wager {\em et~al.}(2014)Wager, Fithian, Wang, and
  Liang]{Wager2014Altitude}
Wager, S., Fithian, W., Wang, S., and Liang, P.~S. (2014).
\newblock Altitude training: Strong bounds for single-layer dropout.
\newblock In Z.~Ghahramani, M.~Welling, C.~Cortes, N.~D. Lawrence, and K.~Q.
  Weinberger, editors, {\em Adv. Neural. Inform. Process Syst.}, pages
  100--108. Curran Associates, Inc.

\bibitem[Zeisel {\em et~al.}(2015)Zeisel, Manchado, Codeluppi, Lonnerberg,
  La~Manno, Jureus, Marques, Munguba, He, Betsholtz, Rolny, Castelo-Branco,
  Hjerling-Leffler, and Linnarsson]{Zeisel2015Cell}
Zeisel, A., Manchado, A. B.~M., Codeluppi, S., Lonnerberg, P., La~Manno, G.,
  Jureus, A., Marques, S., Munguba, H., He, L., Betsholtz, C., Rolny, C.,
  Castelo-Branco, G., Hjerling-Leffler, J., and Linnarsson, S. (2015).
\newblock {Cell types in the mouse cortex and hippocampus revealed by
  single-cell RNA-seq}.
\newblock {\em Science\/}, {\bf 347}(6226), 1138--42.

\bibitem[Zou and Hastie(2005)Zou and Hastie]{Zou2005regularization}
Zou, H. and Hastie, T. (2005).
\newblock Regularization and variable selection via the {E}lastic {N}et.
\newblock {\em J. R. Stat. Soc. Ser. B\/}, {\bf 67}, 301--320.

\end{thebibliography}

\bibliographystyle{natbib}

\end{document}